# Full-Color Computational Imaging with Single-Pixel Detectors Based on a 2D Discrete Cosine Transform

Bao-Lei Liu,[1] Zhao-Hua Yang,[1,*] Ling-An Wu[2]
[1]*School of Instrument Science and Optoelectronics Engineering, Beihang University, Beijing 100191, China*
[2] *Institute of Physics and Beijing National Laboratory for Condensed Matter Physics, Chinese Academy of Sciences, Beijing 100190, China*
**Corresponding author:* yangzh@buaa.edu.cn*

**We propose and demonstrate a computational imaging technique that uses structured illumination based on a two-dimensional discrete cosine transform to perform imaging with a single-pixel detector. A scene is illuminated by a projector with two sets of orthogonal patterns, then by applying an inverse cosine transform to the spectra obtained from the single-pixel detector a full-color image is retrieved. This technique can retrieve an image from sub-Nyquist measurements, and the background noise is easily canceled to give excellent image quality. Moreover, the experimental setup is very simple.**

## 1. INTRODUCTION

Ghost imaging (GI), a nonlocal imaging method whereby an object is reconstructed by means of intensity correlation between the object beam and a corresponding reference beam has attracted considerable attention in recent years. The first such experiment was performed using entangled photon pairs obtained by spontaneous parametric downconversion [1]. Later, GI was realized using pseudothermal light [2-5] and true thermal light [6-8]. Though the light sources may be different, conventional GI requires a beam splitter to produce correlated light fields in two separate beams, one beam passing via the object to a bucket detector collecting only the total light intensiy, and the other measured directly by a detector with spatial resolution.

Computational ghost imaging, proposed in 2008 [9], eliminated the need for a beam splitter by using a spatial light modulator capable of creating deterministic speckle patterns to illuminate the object [10, 11]. This system is more practical and suitable for remote sensing since the high resolution detector in the reference beam is replaced by a computer generated propagating field. In GI a ghost image of the object is retrieved by correlating the object beam with the reference beam, but in computational GI the measurements recorded by the bucket (single-pixel) detector in the object beam are convoluted with precomputed intensity distribution patterns.

When combined with compressive sensing (also known as compressive sampling) [12-14], ghost imaging can reconstruct an image from the data sampled at sub-Nyquist frequencies, and can retrieve an image consisting of N2 pixels using much fewer than N2 measurements with a single-pixel detector [15]. Compressed sensing exploits the redundancy in the structure of most natural objects to reduce the number of measurements required, and is widely applied in the field of computational imaging.

Recently, a new computational imaging technique was proposed by Zhang et al, which can reconstruct high-quality images by acquiring their Fourier spectrum [16]. Employing an approach called four-step phase-shifting sinusoidal illumination, and after illuminating a scene with four two-dimensional (2D) sinusoid patterns, every coefficient of the scene's Fourier spectrum is acquired by four responses. The image is then reconstructed by applying an inverse Fourier transform. This is a compressive sampling-like approach, since most natural images are sparse in the Fourier domain. For image reconstruction, only the light intensities collected by the bucket detector need to be considered, without the need to correlate with the precomputed illumination patterns, which are recovered in the inverse Fourier processing. In this paper, we propose a different full-color computational imaging method using single-pixel detectors based on a 2D discrete cosine transform (DCT), which is simpler and requires fewer measurements. The spectral coefficients can be obtained after illuminating the scene with only two groups of orthogonal sinusoidal patterns, allowing each coefficient to be acquired from just two responses. The image can then be retrieved by applying an inverse cosine transform to the spectrum acquired. Since this transform provides the same features of energy compaction as a discrete Fourier transform for most natural images [17], our technique is also a compressive sampling-like approach, which means an image can be reconstructed from only a few coefficients of the spectrum collected by the single-pixel detector.

## 2. THEORY

The object or target scene is illuminated by a projector with two sets of orthogonal sinusoidal patterns, and the reflected field intensity is collected by a single-pixel detector. Every coefficient of the image's cosine transform spectrum is acquired from the two detector responses corresponding to the two patterns. The scene is retrieved by applying the inverse DCT algorithm to the cosine transform spectrum. The final image's signal-to-noise ratio is very good, since any unwanted background illumination is automatically canceled, as will be shown below.

The cosine transform employed by Ahmed et al. [18] expresses a finite sequence of data points in terms of a sum of cosine functions of different frequencies, and has found wide application in transform image coding. It is the foundation of the JPEG standard for still-image coding and the MPEG standard for moving images.

The 2D DCT and inverse DCT of an image array are defined in series form as [19]

$$F(u,v) = \frac{2}{\sqrt{MN}} \sum_{x=0}^{M-1} \sum_{y=0}^{N-1} C(u)C(v) f(x,y) \cos\frac{(2x+1)u\pi}{2M} \cos\frac{(2y+1)v\pi}{2N}, \quad (1)$$

$$I(x,y) = \frac{2}{\sqrt{MN}} \sum_{u=1}^{M-1} \sum_{v=1}^{N-1} C(u)C(v) F(u,v) \cos\frac{(2x+1)u\pi}{2M} \cos\frac{(2y+1)v\pi}{2N}. \quad (2)$$

Here, $C(0) = (2)^{-1/2}$ and $C(w) = 1$ for $w = 1, 2, \ldots, N-1$, and $f(x,y)$ is the spatial distribution of the original image with $M \times N$ pixel resolution, which also represents the image of the target scene as described in the latter part of this paper; $F(u,v)$ is the frequency spectrum of the transformation where $u$ and $v$ are the horizontal and vertical spatial frequencies, respectively. The reconstructed image $I(x,y)$ is obtained by applying the inverse DCT.

We simplify Eqs. (1) and (2) into the two equations below in order to describe our technique more effectively:

$$F(u,v) = \mathcal{DCT} \times f(x,y) \quad (3)$$
$$I(x,y) = \mathcal{IDCT} \times F(u,v) \quad (4)$$

where $\mathcal{DCT}$ represents the discrete cosine transform and $\mathcal{IDCT}$ is the inverse transform. One of the key steps of our technique is to realize the $\mathcal{IDCT}$ physically to obtain the spectral components. However, the $\mathcal{DCT}$ in the above equation is a

2D orthogonal product of two cosine functions and contains a considerable amount of negative numbers. Therefore, we add a constant term $a$ to both sides, equal to the average intensity of the image. This makes all elements natural numbers, as shown by Eqs. (5) and (6), where $b$ represents the contrast. Then, the DCT coefficients of the object can be obtained by Eq. (7). The background illumination $e$ that exists in the experiment is added to Eqs. (5) and (6) as a constant, so the constant $a$ and the background illumination value $b$ are effectively canceled out. We thus have

$$F_1(u,v) = (a + b \times DCT) \times f(x,y) + e, \qquad (5)$$

$$F_2(u,v) = (a - b \times DCT) \times f(x,y) + e, \qquad (6)$$

$$F(u,v) = (F_1(u,v) - F_2(u,v)) / (2 \times b). \qquad (7)$$

The multiplier of $f(x,y)$ in Eqs. (5) and (6) can be expressed as the Eqs. (8) and (9) below, respectively:

$$P_1(u,v) = a + b \frac{2}{\sqrt{MN}} C(u) C(v) \sum_{x=0}^{M-1} \sum_{y=0}^{N-1} \cos\frac{(2x+1)u\pi}{2M} \cos\frac{(2y+1)v\pi}{2N}, \qquad (8)$$

$$P_2(u,v) = a - b \frac{2}{\sqrt{MN}} C(u) C(v) \sum_{x=0}^{M-1} \sum_{y=0}^{N-1} \cos\frac{(2x+1)u\pi}{2M} \cos\frac{(2y+1)v\pi}{2N}. \qquad (9)$$

where $P_1(u,v)$ and $P_2(u,v)$ are two patterns with the spatial frequencies $(u,v)$. Figure 1(a) and (b) show part of the two sets of the illumination patterns generated from Eqs. (8) and (9), respectively. The patterns are a combination of horizontal and vertical frequencies for a 64×64 ($M = N = 64$) 2D DCT. Each step from left to right and top to bottom is an increase in frequency of 1/2 cycle. For example, moving right (or down) one step from the top-left square yields a half-cycle increase in the horizontal (or vertical) frequency. The source data (64×64) is transformed to a linear combination of these 64 horizontal and vertical frequencies.

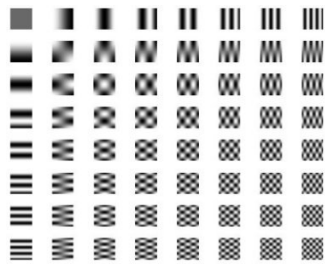

(a)

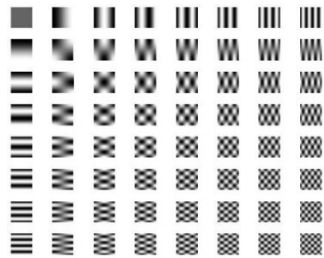

(b)

**Figure 1.** Part of the two sets of illumination patterns for $M = N = 64$ used in this paper. (a) Patterns generated from Eq. (5). (b) Patterns generated from Eq. (6).

By employing the inverse 2D DCT algorithm of Eq. (4), the original image can be reconstructed by the computer.

The imaging process of our proposed system can be divided into five modules, as shown in figure 2. A projector directs the patterns generated by a computer onto the target scene, and the single-pixel detector measures the reflected light from the target scene in synchronization and transfers it to an image reconstructor. The two sets of field intensity measurements at the detector, $D_1(u,v)$ and $D_2(u,v)$, both have spatial frequencies of $(u,v)$. Since the multiplication of $F(u,v)$ by a constant makes no difference to the reconstruction of the image by the inverse DCT, $F(u,v)$ can be calculated from Eq. (10). The image of the target scene can then be reconstructed from Eq. (2) by substituting.

$$F(u,v) = D_1(u,v) - D_2(u,v) \qquad (10)$$

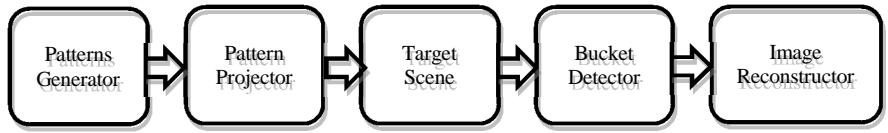

**Figure 2.** Procedure of computational imaging with single-pixel detectors based on 2D DCT.

Since DCT has a strong "energy comspaction" property [17, 18], most of the signal information tends to be concentrated in a few low-frequency components, allowing the small high-frequency components to be discarded. Thus the technique proposed here is somewhat similar to a compressive sampling approach in that the image can be reconstructed with only a fraction of the spectra.

## 3. EXPERIMENTAL SETUP

The setup for this experiment is shown in figure 3. A digital light projector (BENQ MS513P) provides sinusoidal orthogonally structured patterns to illuminate a 3D scene consisting of a flower in front of a background picture $12 \times 12$ cm in size at a distance of about 120 cm. The main components of the projector are a digital micro-mirror array, a high voltage mercury lamp, and a lens of 55 mm focal length. The light reflected from the scene is directed onto a CMOS sensor with a Bayer filter (Basler acA1920-155uc) which detects red (R), green (G) and blue (B) with different filters at different pixels. The signals from all the color channels are recorded in the computer, and the full-color image is reconstructed by the algorithm shown above.

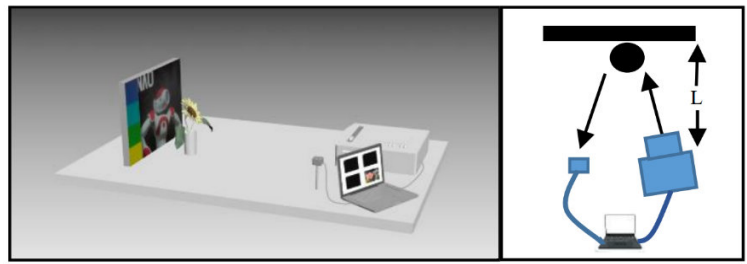

**Figure 3.** Experimental system for imaging a colored 3D scene. The projector illuminates the scene with two sets of orthogonal patterns. The distance L between the projector and target scene is about 120 cm.

To synchronize the system we designed the data collecting program written in C++ language using Microsoft Visual Studio, so that we could project the patterns and record the photodetector signals synchronously. The exposure time of each image was 0.1 s, and the patterns were switched at 8 frames per second.

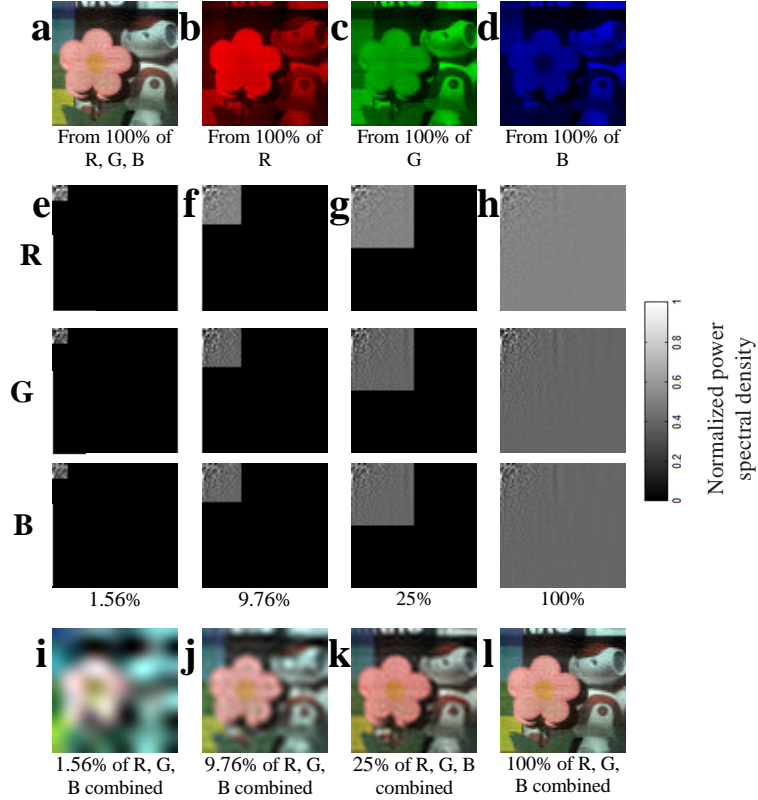

**Figure 4.** (a) Full-color reconstructed image of the scene, obtained by combining the separate reconstructions from the (b) red, (c) green and (d) blue spectral data. (e)-(h) are the density distributions of the cosine transform spectra acquired for the three colors with different spectral coverages of 1.56, 9.76, 25, and 100%, with each row corresponding to red, green and blue, respectively. (i) – (l) are the reconstructed combined R, G, B images with the same spectral coverages of (e)-(h), respectively.

## 4. IMAGE RECONSTRUCTION

Using two groups of orthogonal sinusoidal patterns, the scene's illumination field was changed a total of $64 \times 64 \times 2$ times, for the case of $M = N = 64$. The integrated intensities of the reflected field were measured $64 \times 64 \times 2$ times, accordingly. The spectrum of each image can be calculated quickly by Eq. (8). The size of the illumination patterns and final reconstruction was $64 \times 64$ pixels. The red, green and blue images of the scene are reconstructed by the algorithm based on the 2D DCT, as described above. The entire full-color image with all the colors combined is shown in figure 4 (a). Figures 4 (b) - (d) show the red, green and blue reconstructions, respectively. It should be noted that the results retrieved from any one of the three colors are graded in tone.

Figures 3 (e) - (h) are the density distributions of the cosine transform spectra for the three colors acquired with different spectral coverages of 1.56, 9.76, 25, and 100%, respectively. Since the scene image is sparse in the frequency domain, the the image energy spectra, as we can see, is concentrated in the top left-hand corner of the full spectra, where the low-frequency components are located. Figure 3 (i) - (l) are the reconstructed images with the same spectral coverage of red, green and blue corresponding to figure 3 (e) - (h). It is evident that the quality improves as the covering spectrum increases. The image is already recognizable when the coverage reaches 9.76%, which means that an image with $64 \times 64$ pixel resolution can be retrieved from $400 \times 2$ measurements by our technique.

## 5. SUMMARY

In conclusion, this paper demonstrates a full-color computational imaging technique based on 2D DCT with single-pixel detectors, which can be used to produce a full-color image of a 3D scene. Compared with conventional, computational or compressive sensing GI, this method retrieves an image directly from the total light intensity of the illuminated scene, instead of having to correlate an object beam and reference beam. The number of measurements can be greatly reduced, and the image quality is very good, while the experimental setup and data processing are simple and low-cost. Furthermore, with this method it should be possible to retrieve an image even when the light to be collected is distorted by a scattering medium. Future applications could include remote sensing using a laser modulated by a spatial light modulator to illuminate the target scene.

**Funding.** National Natural Science Foundation of China (Grant Nos. 61473022 and 60907031)